\documentclass[prd,aps,twocolumn,floatfix]{revtex4}

\usepackage[pdftex]{hyperref}

\usepackage{amssymb,graphicx}
\usepackage{epsfig}
\usepackage[usenames,dvipsnames]{color}

\def\apjl{Ap. J. Letters}

\def\physrep{Physics Reports}

\def\Alfven{Alfv{\'e}n\, }

\newcommand{\had}{{\sc had}}

\begin{document}

\title{Boosting jet power in black hole spacetimes}

\author{David Neilsen$^{1,2}$,
Luis Lehner$^{2,3,4}$,
Carlos Palenzuela$^{5,6}$,\\
Eric W. Hirschmann$^{1}$,
Steven L. Liebling$^{7}$,
Patrick M. Motl$^{8}$,
Travis Garrett$^{2,6}$}
\affiliation{
$^{1}$Department of Physics and Astronomy, Brigham Young University, Provo, UT 84602, USA, \\
$^{2}$Perimeter Institute for Theoretical Physics,Waterloo, Ontario N2L 2Y5, Canada,\\
$^{3}$Department of Physics, University of Guelph, Guelph, Ontario N1G 2W1, Canada, \\
$^{4}$CIFAR, Cosmology \& Gravity Program, Canada, \\
$^{5}$Canadian Institute for Theoretical Astrophysics, Toronto, Ontario M5S 3H8, Canada, \\
$^{6}$Department of Physics \& Astronomy, Louisiana State University, Baton Rouge, LA 70802, USA, \\
$^{7}$Department of Physics, Long Island University, New York 11548, USA, \\
$^{8}$Department of Science, Mathematics and Informatics, Indiana University Kokomo,
             Kokomo, IN 46904-9003, USA}

\date{\today}

%
%
\begin{abstract}
The extraction of rotational energy from a spinning black hole via the Blandford-Znajek mechanism
has long been understood as an important component in models to explain  energetic jets from compact astrophysical sources.
Here we show more generally that the kinetic energy of the black hole,
both rotational and translational, can be tapped,
thereby
producing even more luminous
jets powered by the interaction of the black hole with its surrounding plasma.
We study the resulting Poynting jet
that arises from single boosted  black holes and binary black hole systems.
In the latter case, we find that increasing the orbital angular
momenta of the system and/or the spins of the individual black holes results 
in an enhanced Poynting flux.
\end{abstract}

\maketitle

%
%
\section{Introduction}
Enormously powerful events illuminate the universe that
challenge our understanding of the cosmos. Indeed, intense electromagnetic
emissions of order
$\gtrsim 10^{51}{\rm ergs}$~\cite{Hamuy:2003xv,Woosley:2006fn,2001ApJ...562L.149M} have
 been routinely observed in supernovae, gamma ray bursts~(GRBs),
and active galactic nuclei~(AGNs), for example.
However, despite important theoretical and observational
advances, we still lack a thorough understanding of these systems (e.g.~\cite{Fender:2010tk}).
Intense observational and theoretical efforts are
ongoing in order
to unravel these fascinating phenomena. While the full details
remain elusive, one of the
natural ingredients in theoretical models is the inclusion of
a rotating black hole which serves to
convert binding and rotational energy of the system to electromagnetic 
radiation
in a highly efficient manner.
The starting point for these theoretical models can be traced back to ideas laid out by
Penrose~\cite{Penrose:1969pc} and Blandford and Znajek~(BZ)~\cite{Blandford:1977ds},
which explain the extraction of energy from a rotating black hole. These seminal studies, along
with subsequent work (see references in e.g.~\cite{2000PhR...325...83L,2002luml.conf..381B,2008ASSL..355.....P}),
have provided a basic understanding of highly energetic emissions from single black hole systems
interacting with their surroundings.

Recent work has
indicated that
related
systems
can also tap kinetic energy and lead to powerful jets~\cite{sciencejets}.
This work concentrated
on non-spinning black holes moving through a plasma and highlighted that relative black hole motion
alone (with respect to a stationary electromagnetic field topology at far distances from it) can induce the
production of jets. Furthermore, subsequent work~\cite{prdjets} demonstrated that even black holes with misaligned spins
with respect to the asymptotic magnetic field direction induce strong emissions with power
comparable to the aligned case. 
These studies suggested that, independent of their inclination, astrophysical jets might be powered  by the efficient
extraction of both rotational and translational kinetic energy of black holes,
and inducing even more powerful jets than the standard BZ mechanism 
would suggest.

Galactic mergers provide a likely scenario for the production of the 
binary black hole 
systems considered here~\cite{1980Natur.287..307B,Milosavljevic:2004cg}.
In such a merger, the supermassive
black holes associated with each galaxy will ultimately form a binary in the merged galaxy.
A variety of
interactions will tighten
the black hole binary.
Eventually
the dynamics of the system will be
governed by gravitational radiation reaction which drives the binary to merge.
The circumbinary disk will likely be magnetized
and thereby anchor magnetic field lines,
some of which will traverse the central region containing the binary.
Preliminary observational evidence for supermassive black hole
binaries resulting from galactic mergers has already been presented~\cite{Comerford:2008gm}.

An ambient magnetic field threading a spinning black hole populates a low
density plasma surrounding the black hole as explained by BZ~\cite{Blandford:1977ds}.
Even for black holes with no spin, it was recently shown that the
orbiting binary interacting with the surrounding plasma
can lead to a collimated Poynting flux~\cite{sciencejets}.
In this work, we consider this
basic paradigm of energy extraction from black holes
with the additional complexity of intrinsic black hole spin.
Binaries consisting of spinning black holes demonstrate 
similar, albeit energetically enhanced, phenomenology. 
Furthermore, we
investigate the dependence of the energy flux on the black hole velocity
and highlight a resulting strong boost in the emitted power.

In addition to exquisite and powerful
electromagnetic detectors, soon gravitational waves will be
added to the arsenal of phenomena employed to understand our cosmos.
These
studies suggest excellent prospects 
for the coincident detection of
both electromagnetic and gravitational signals
from binary black hole systems. 
Certainly, dual detection of electromagnetic and gravitational wave signals
would transform  our understanding of these systems and lead to the refinement of
theoretical models (e.g~\cite{2002luml.conf..207S,2003ApJ...591.1152S,2009arXiv0902.1527B,2009astro2010S.235P}).

In the remainder of this paper, we
elucidate the basic phenomenology arising from the
interaction of binary black hole
systems
immersed
within a plasma environment.
We
focus in particular
on understanding
the Poynting flux emissions from such binaries as well as single, possibly spinning,
black holes moving through plasma.
We describe our
equations and assumptions employed together with some of the details of
our numerical implementation in Sec.~\ref{sec:implementation}.
Sec.~\ref{sec:results} describes our results
for both single and binary black holes and we provide concluding
comments in Sec.~\ref{sec:conclusions}.

%
%
\section{Implementation details}
\label{sec:implementation}
The combined gravitational and electromagnetic systems that we consider
consist of black hole spacetimes in which the black holes can be regarded
as immersed in an external magnetic field. Such fields, as mentioned above, will be
anchored to a disk.  We consider this disk to be outside our computational domain
but its influence is realized through the imposition of
suitable boundary conditions on the corresponding electric and magnetic fields
on the boundaries of our domain~\cite{sciencejets,prdjets}.
With regard to the
magnetosphere around the black holes, we assume that
the energy density
of the magnetic field dominates over its tenuous density such that the
inertia of the plasma can be
neglected. The magnetosphere is therefore treated within
the force-free approximation~\cite{1969ApJ...157..869G,Blandford:1977ds}.
We note that the contribution of the energy density of the plasma
to the dynamics of the spacetime is negligible and we can ignore
its back reaction on the spacetime.

We use the BSSN
formulation~\cite{Baumgarte:1998te,Shibata:1995we} of the
Einstein equations and
the force-free equations as described in~\cite{sciencejets,prdjets}.
We discretize the equations using finite difference techniques on a 
regular Cartesian grid and use adaptive mesh refinement (AMR)
to ensure that sufficient resolution is available where 
required in an efficient manner.  We use the \had\ computational 
infrastructure,
which provides distributed, Berger-Oliger
style AMR~\cite{had_webpage,Liebling} with full sub-cycling
in time, together with an improved treatment of artificial 
boundaries~\cite{Lehner:2005vc}.
Refinement regions are determined using truncation error estimation provided by a shadow
hierarchy~\cite{pretorius}, which adapts dynamically to ensure the 
estimated error is bounded by a
pre-specified tolerance. Typically our adopted values result in a grid hierarchy
yielding a resolution such that $40$ grid points in each direction cover each black hole.
We use a fourth order accurate spatial discretization 
and a third order accurate in time Runge-Kutta integration scheme~\cite{binaryns}. 
We adopt a Courant
parameter of $\lambda = 0.4$ so that $\Delta t_\ell = 0.4 \Delta x_\ell$
on each refinement level $\ell$.
In tests performed here for the coupled system 
(and in our previous works for the force-free Maxwell equations), 
the code demonstrates convergence while maintaining small constraint
residuals for orbiting black holes. Furthermore, we obtain for orbiting black
hole evolutions agreemend with runs from other codes for the same initial data.

To extract physical information, we monitor the
Newman-Penrose electromagnetic ($\Phi_2$) and gravitational ($\Psi_4$) radiative
scalars \cite{Newman:1961qr}.
These scalars are computed by contracting the Maxwell and the Weyl tensors respectively,
with a suitably defined null tetrad (as discussed in~\cite{Lehner:2007ip}),
\begin{eqnarray}
  \Phi_2 = F_{ab} n^a \bar m^b \, , \quad
  \Psi_4 = C_{abcd} n^a \bar m^b n^c \bar m^d\, ,
\end{eqnarray}
and
they
allow us to
account for the energy carried off by outgoing waves at infinity.
The scalar $\Phi_2$ (essentially the radial component
of the Poynting vector) provides a measure of the electromagnetic radiation
at large distances from an isolated system. However, as the system studied here
has a ubiquitous magnetic field, special care must be taken to compute the energy flux.
We account for this difficulty by subtracting the scalar
$\Phi_0 = -F_{ab} l^a m^b$ from $\Phi_2$.
Hence, from here forward, by $\Phi_2$ we mean the difference $\Phi_2 \rightarrow \Phi_2 - \Phi_0$.
The
luminosities in
electromagnetic and gravitational waves are given by the integrals of the fluxes
\begin{eqnarray}
  L_{EM} &=& \frac{{dE}^{EM}}{dt} =
     \lim_{r \rightarrow \infty}  \int r^2 |\Phi_2|^2 d\Omega ~,
\label{FEM} \\
  L_{GW} &=& \frac{{dE}^{GW}}{dt\,} =
     \lim_{r \rightarrow \infty} \int \frac{r^2}{16 \pi} \left| \int_{\infty}^t \Psi_4 dt' \right|^2 d\Omega~.
\label{FGW}
\end{eqnarray}

We assume that the black holes are immersed in an initially constant
magnetic field,
such as one produced by a distant disk
surrounding the black hole system. For the binary systems we consider,
the orbital plane of the evolution is assumed to be aligned
with that of the circumbinary disk. The magnetic field is anchored
in the disk with its associated magnetic dipole aligned with the orbital
angular momentum (chosen along the $\hat z$ direction).
Initially, the magnetic field is set
to be perpendicular to the velocity of the black holes and the electric
field is set to zero.
Because the electromagnetic field is affected by the spacetime curvature,
it will be dynamically distorted from its initial configuration and
generate a transient
burst of both gravitational and electromagnetic waves
as it settles into a physically relevant and dynamical configuration.

The initial magnitude of the magnetic field $B_0$ is chosen to be
consistent with astrophysically relevant
values~\cite{2008A&A...477....1M,1993ApJ...403...94F}.
We present our results for a field strength which is bounded by the
the Eddington magnetic field strength $B\simeq 6 \times 10^{4}
(M/10^8\,M_{\odot})^{-1/2}$~G~\cite{2008arXiv0810.1055D}.
As noted above, for these values the plasma's energy is several orders of
magnitude smaller than that of the gravitational field.  Thus, although the
plasma is profoundly affected by the black holes, it
has a negligible influence on their dynamics.

%
%
\section{Results}
\label{sec:results}

Here we discuss our results for both boosted and binary black hole systems.
For simplicity we introduce the notation $L_{43} = 10^{43}$~ergs/s and compute
dimensionful quantities with respect to a representative system with total mass
$10^8 M_{\odot}$ immersed in a magnetic field with strength $10^4$~G. We explicitly
provide proportionality factors for calculations for other configurations.
Quantities in geometric units are calculated by setting $G=c=1$.\\

\noindent {\em Single black holes}

For single, boosted black holes, we adopt a computational
domain defined by $[-320\,m_b,320\,m_b]^3$ and assume a fixed ``bare"
mass
of the black hole (the mass it would have if it were static and in isolation)
of $m_b=1$ in geometric units.
We set the initial linear momentum of the black hole to
${\vec p} = 0.05\,\, n\, {\hat i}$ (with $n$ an integer that we vary between $0$
and $4$),
and we set the intrinsic angular momentum parameter
$\vec a \equiv \vec J/M^2$ to be $0$ or $0.6 \, \hat k$.
For these parameters we study the resulting electromagnetic collimated flux energy
and its dependence on both boost velocity and black hole spin.

\begin{figure}
$\begin{array}{c@{\hspace{0.1in}}c@{\hspace{0.1in}}c}
\includegraphics[width=1.7in]{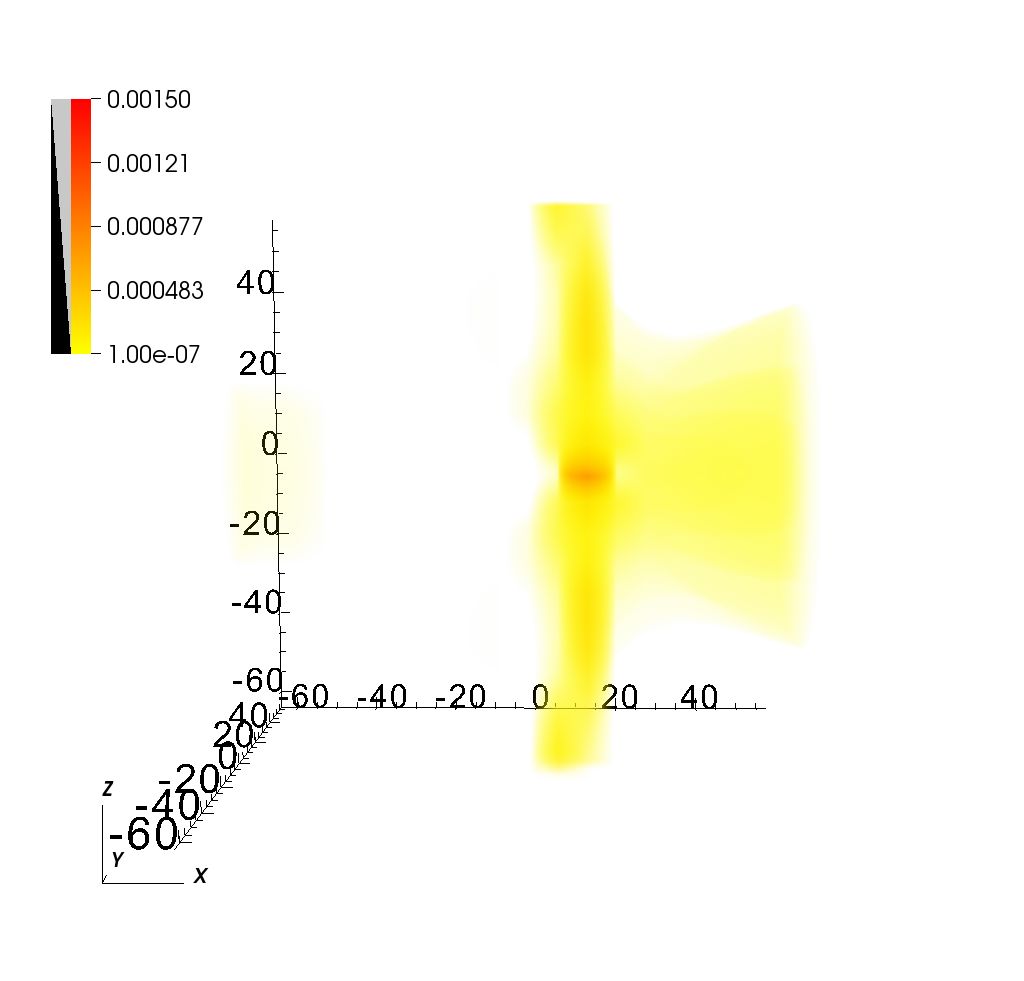} &
\includegraphics[width=1.7in]{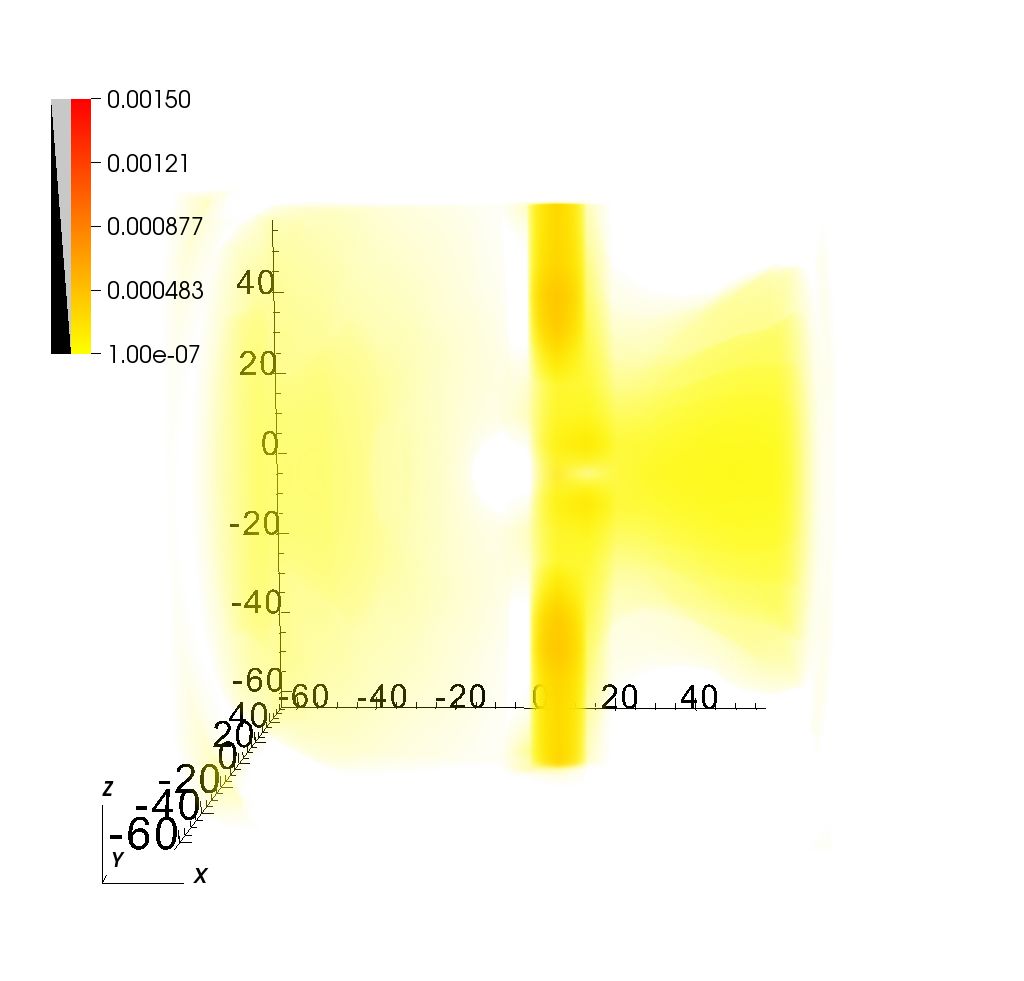}
\end{array}$
\caption{
Zoom-in of the electromagnetic emission $|r \Phi_2|^2$
from a boosted black hole (with $\vec p = 0.15 \hat x$). The
left frame shows the non-spinning black hole case ($a=0$), while the right frame shows
the spinning case ($a = 0.6$). The color scale is the same for both frames and the black hole
is boosted to the right. The stronger emission in the vertically collimated region of the
right frame is apparent.
}
\label{fig:single_boost3D}
\end{figure}

Fig.~\ref{fig:single_boost3D} illustrates the qualitative features of the
electromagnetic emission for a boosted black hole with and without spin.
A collimated emission is clearly induced along the asymptotic magnetic
field direction.
Significantly, an even stronger emission is obtained for the spinning case as expected.

\begin{figure}[h]
\begin{center}
\includegraphics[width=4.6cm]{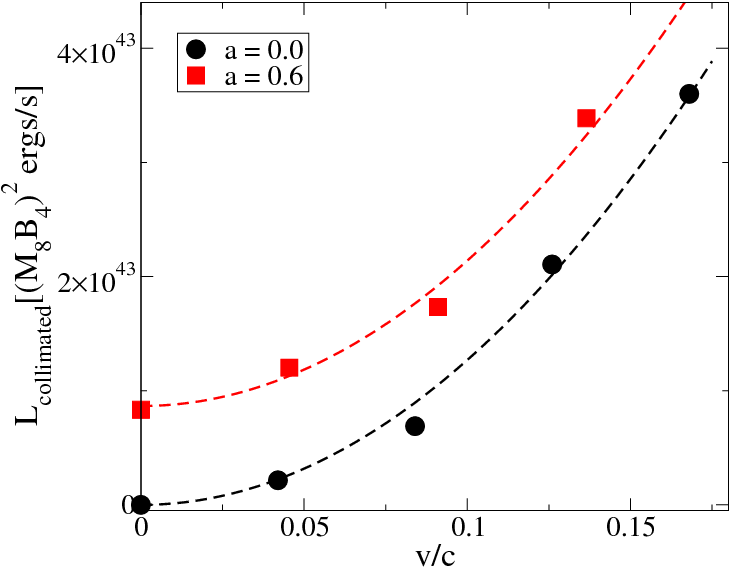}
\end{center}
\caption{Collimated luminosity as a function of the boost velocity of the black
hole for two different spin magnitudes.
The dashed lines are obtained
from a quadratic fit of the form  $L_{\rm collimated} = L_{\rm spin} + L_2 v^2$
where $L_{\rm spin}$ and $L_2$ are two constants. The constant $L_2$ is the same for both
fits, supporting the argument that the luminosity does contain two different components, one for spin and one for boost.
}
\label{fig:single_boost}
\end{figure}

To achieve a more quantitative understanding, we
compute a measure of the collimated  electromagnetic luminosity, $L_{\rm collimated}$, by 
integrating the flux over a $15^\circ$ cone of 
a spherical surface centered along the moving black hole
with a
radius $r=20\,m_b$ (roughly equivalent to $40\, R_H$, where $R_H$ is the Schwarzschild
radius of the black hole).  We present our results with respect to $v$
(the measured {\em coordinate velocity} of the black hole) and
plot the collimated luminosity achieved once the system
reaches a quasi-stationary state (in other words, after the initial transient stage)
for the spinning and non-spinning cases
in Fig.~\ref{fig:single_boost}.

Several key observations are evident from the figure.
For $v=0$, the electromagnetic
energy luminosity does not vanish  
for the spinning black hole, while
it does vanish for the non-spinning black hole.
This is expected, as the spinning black hole interacts with a surrounding plasma
and radiates by the Blandford-Znajek
mechanism~\cite{Blandford:1977ds}. This luminosity results from the plasma's ability to
extract rotational energy from the black hole and power a jet with an energy luminosity scaling
as $L_{BZ} \approx \Omega_H^2 B^2$ ~\cite{2010ApJ...711...50T,prdjets} (with
$\Omega_H \equiv a/(2\, R_{H})$ the rotation frequency associated with the
black hole).

For $v \neq 0$, both the spinning and non-spinning black holes
have a non-trivial associated energy flux.
In the latter case this flux
arises solely
from the ability of the system to tap translational kinetic energy from the black hole, while the former
results from the extraction of both translational and rotational kinetic energies.

The non-spinning black hole shows the expected $v^2$ dependence in
the electromagnetic energy luminosity (see Fig.~\ref{fig:single_boost})
consistent with the membrane picture of a black hole as a conductor.
Indeed, as was already indicated in the spinless case~\cite{Palenzuela:2009yr,Palenzuela:2009hx,sciencejets,prdjets},
a black hole moving (with speed $v$) through an ambient magnetic field acquires an induced
charge separation ($\propto v$).
The membrane paradigm~\cite{1986bhmp.book.....T} of a black hole explains this
induced charge by regarding the black hole as a conductor moving through a magnetic field,
and hence the induced charge is analogous to the classical Hall effect.
With this observation and the induction equation, it is straightforward to conclude an electromagnetic
energy flux will be produced with magnitude $\propto v^2 B^2$. Such a quadratic
dependence on speed is apparent in the figure.
These results are also consistent with the work of~\cite{1965JGR....70.3131D}
which studied the Poynting flux associated with a moving conductor in a magnetized plasma,
in particular the motion of artificial satellites in orbit.
They found that this flux
obeys $L_{v} \approx (v/v_{\rm alf})^2 B^2$~\cite{1965JGR....70.3131D} where
$v_{\rm alf}$, the propagation speed
of the \Alfven modes,
is the speed of light in a force-free environment.
Furthermore, a misalignment is expected between the collimated energy flux and the original magnetic
field orientation such that $\tan(\alpha) = v/v_{\rm alf}$. With the cautionary note that measuring this angle
is a delicate enterprise within General Relativity, especially in the strongly curved
region around
the black hole, such a relation is manifested in our results. For instance, for $a=0$, $p_x=0.10$ the
measured angle obeys $\tan(\alpha)=0.07$ instead of the predicted $0.08$~\cite{1965JGR....70.3131D}.\\

The luminosity for the spinning black hole also demonstrates the same
quadratic dependence in velocity, with the additional component
from the spin.  Notice
that the difference between the obtained luminosities 
in the spinning and spinless
cases remains fairly
constant for the different values of $v$. Thus, to a reasonably good approximation,
we can express the luminosity obtained as a sum of 
a spin dependent component and
a speed dependent one, $L_{\rm collimated} \simeq L_{\rm spin} + L_{\rm speed}$, the
latter scaling as $L_{\rm speed}=L_2 v^2$ .
A fit to the case $a=0$ gives $L_2 = 127 \, \left(M_8\, B_4\right)^2 L_{43}$. Next, using
this value for the case $a=0.6$, we obtain the fit for $L_{\rm spin}  =     L_1 =0.87 \, \left(M_8\, B_4\right)^2 L_{43}$.
We can then provide an estimate
of the associated luminosity for general cases as follows.

The spin component for the general case can be expressed in terms of its dependence on
the the black hole spin  as
$L_{\rm spin} = L_1 (\Omega_H(a)/\Omega_H(a=0.6))^2$.
The boost component takes the already mentioned quadratic form $L_{\rm speed} = L_2 v^2 $.

Therefore, we have for the estimate
\begin{eqnarray}
L_{\rm collimated} & = & L_{\rm spin} + L_{\rm speed} \, , \cr 
                   & = & L_1 \,     \left(\frac{\Omega_H |_{a}}{\Omega_H |_{0.6}}\right)^2 + L_2 \, v^2 \, , \cr
                   & = & \left[0.87 \left(\frac{\Omega_H |_{a}}{\Omega_H |_{0.6}}\right)^2 + 127 v^2 \right] L_{43}.
\end{eqnarray}
Notice that for $a=0.6$, the non-rotational contribution
$L_{\rm speed}$ to the emitted power becomes larger than the rotational one for speeds
approximately
$v>0.08c$.
This relationship, for example, would predict a luminosity for an $a=0.95$, $v=0.5 c$ black
hole to be
$\simeq 36\, L_{43} (M_8\, B_4)^2 $.
This phenomenology strongly suggests that the
Poynting flux can tap both rotational and translational kinetic energies
from the black hole and that faster and more rapidly spinning black holes have
a stronger associated power output.\\

%
%
\noindent {\em Binary black holes}\\
We turn our attention now to the orbiting binary black hole case.
We consider equal-mass configurations in which each
bare black hole has mass $m_b=0.483$ and
the pair is separated by a distance $D \approx 16\, m_b$. We adopt a computational domain of
$[-638\, m_b, 638\, m_b]^3$ and consider 
configurations 
with individual spins either
vanishing or in the up ($+z$) or down ($-z$) orientation
with spin magnitude $|a|=0.515$.
These configurations are summarized in Table~\ref{table:bbh}.

\begin{table}
\begin{center}
    \begin{tabular}{| l | c | l | l | l| l| l ||}
    \hline
    Case & $(p_x, p_y)$ & $a_i$ & $M_{ADM}$ & $J_{ADM}$  \\ \hline
    \hline
    0/0  & $(0.00208,-0.11235)$ & 0 & 1.0 & 0.108 \\ \hline
    u/d  & $(0.00208,-0.11235)$ & $\pm 0.515$ & 1.1 & 0.108  \\ \hline
    u/u  & $(0.00208,-0.11235)$ & $ 0.515$ & 1.1 & 0.349  \\ \hline
    \hline
    \end{tabular}
\end{center}
\caption{Binary black hole configurations considered here. All begin with no momentum
in the $z$ direction.
}
\label{table:bbh}
\end{table}

Before discussing these configurations, we describe our numerical measurements
 which enable a quantitative comparison between the evolutions. First, we compare the
luminosities as functions of gravitational wave frequency, as this is an observable and
allows for a direct comparison of the different cases.
We obtain frequencies from the $l=2,m=2$ gravitational mode.

\begin{figure}[h]
\begin{center}
$\begin{array}{c@{\hspace{0.1in}}c@{\hspace{0.1in}}c}
\includegraphics[width=4.cm]{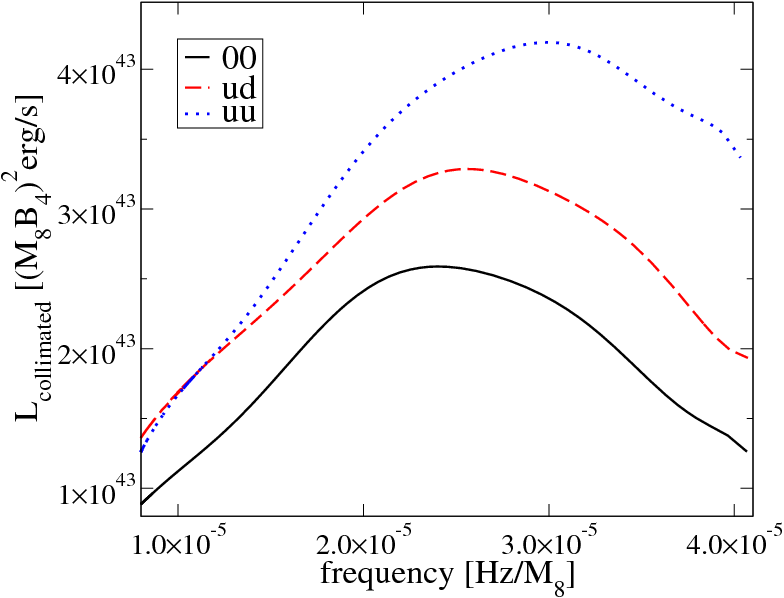} &
\includegraphics[width=4.cm]{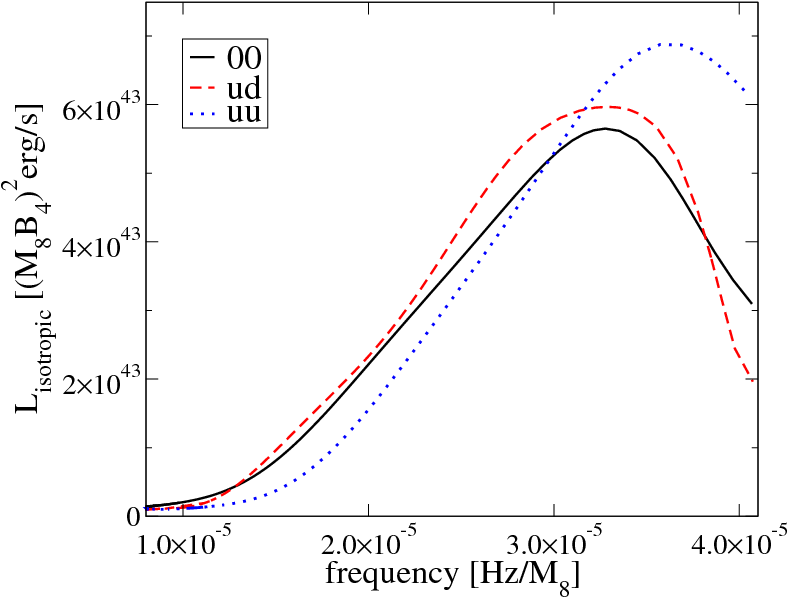}
\end{array}$
\includegraphics[width=4.cm]{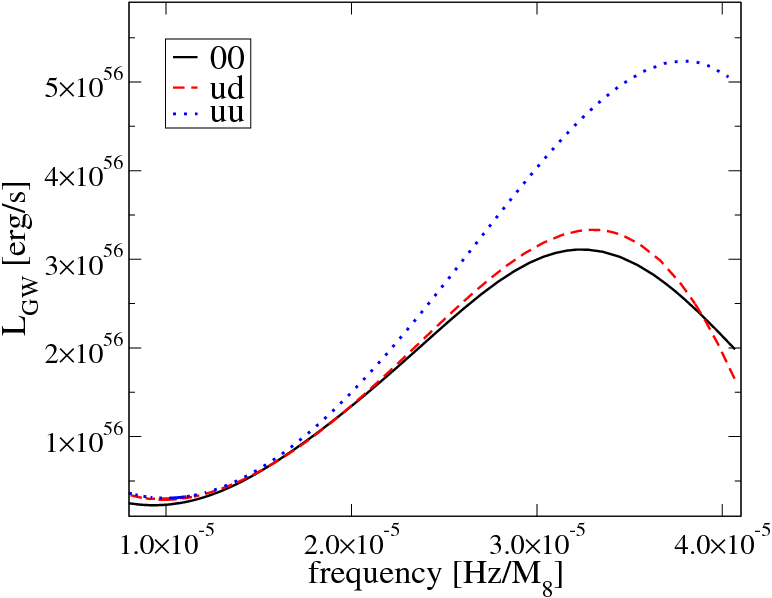}

\end{center}
\caption{Luminosity as a function of the orbital frequency for the binary black
hole configurations described in Table~\ref{table:bbh}, assuming $M=10^8 M_{\odot}$
and $B=10^4$~G. {\bf Top left:} the collimated luminosity associated with the
jets. {\bf Top right:}  the isotropic luminosity representing electromagnetic flux
not associated with the jets. {\bf Bottom:} the gravitational wave output.
}
\label{fig:binary_field}
\end{figure}

Second, we compute three different luminosities for each case:
(i)~the collimated luminosity $L_{\rm collimated}$ obtained by
    integrating the electromagnetic flux over
    a cone of points within $15^\circ$ from the center of mass of the system;
(ii)~the non-collimated, or ``isotropic,'' luminosity  $L_{\rm isotropic}$
     obtained from the integral
     over an encompassing sphere minus the collimated luminosity of (i);
(iii)~the gravitational wave luminosity, $L_{\rm GW}$.
These different luminosities are displayed for the three binary configurations in
Fig.~\ref{fig:binary_field}.

Consider first the 0/0 and u/d configurations which have essentially the same total angular
momentum.
Extensive numerical simulations (see for instance~\cite{Hannam:2009hh}) and simple
estimates~\cite{Buonanno:2007sv} indicate that both binaries will merge into a final black hole
with essentially the same spin ($a\simeq 0.67$), and, so for late times after-merger, the expected
jet structure should be quite similar, dictated by the standard BZ mechanism.

The binary with individual spins aligned reaches a higher
orbital velocity before merger than the previous two cases; thus, the expected maximum power should be higher.
Moreover, the resulting final black hole spins faster ($a \simeq 0.8$)
and thus its BZ associated power will be
higher than that of the previous cases.
Fig.~\ref{fig:binary_field} illustrates this expected behavior by presenting
the Poynting flux energy vs. gravitational wave frequency
for the three cases.

As evident from the figure, at low frequencies where the orbital dynamics
are the same in all cases,
both spinning cases have a higher output than the non-spinning one and the difference
is provided by the spin contribution to the jet emission. Furthermore, both spinning
cases have equal collimated power output because the spin contribution to the luminosity depends only on the
spin magnitude.

We can further examine the basic relation explaining the observed flux by comparing
the three cases studied. We estimate
the black hole {\em coordinate velocities} $v$ and (collimated)
luminosities at three representative frequencies
$\Omega_{i} = \{1; 1.5; 2\} \, 10^{-5}$~[Hz/$M_8$] ($i=1..3$) before the strong non-linear
interaction starts.
Since for these frequencies the measured speeds for the 0/0 and u/d cases are essentially
the same, we notice
that the flux of energy contributed by the spinning black holes can be estimated to be:
\begin{equation}
L^{est}_{BZ} (\Omega) \simeq L_{u/d} (\Omega) - L_{0/0}(\Omega) \, ;
\end{equation}
thus,
\begin{center}
    \begin{tabular}{ | c | c | c|}
    \hline
    $\Omega$ & $(L_{0/0}; L_{u/d}) (M_8 B_4)^2 \, L_{43}$ & $L^{est}_{BZ} (M_8 B_4)^2 \, L_{43}$  \\ \hline
    1   & (1.08; 1.62) &  0.54  \\ \hline
    2 & (1.83; 2.30) &  0.47  \\ \hline
    3  & (2.30; 2.80) &  0.50  \\ \hline
    \end{tabular}
\end{center}
Notice that the $L^{est}_{BZ}$ remains fairly constant through these frequencies which
is evident also in the Fig.~\ref{fig:binary_field}. With this value
one can estimate the luminosity for the u/u case as:
\begin{equation}
L^{est}_{u/u} (\Omega) \simeq L_{0/0}(\Omega) \left(\frac{v_{u/u}(\Omega)}{v_{0/0}(\Omega)}\right)^2  + L^{est}_{BZ}(\Omega) \, .
\end{equation}

Applying such expression to our representative values we obtain,
\begin{center}
    \begin{tabular}{ | c | c | c| c|}
    \hline
    $\Omega$ & $(v_{u/u}; v_{0/0})$ & $(L_{u/u};L^{est}_{u/u}) (M_8 B_4)^2 \, L_{43}$  \\ \hline
    1   & (0.192; 0.174) & (1.64; 1.85)    \\ \hline
    2 & (0.206; 0.186) & (2.52; 2.71)   \\ \hline
    3  & (0.211; 0.187) & (3.26; 3.42)   \\ \hline
    \end{tabular}
\end{center}
thus the estimates are within $\simeq 10\%$ of the measured values while the black holes are
sufficiently separated that their jets do not strongly interact.
Notice that at higher frequencies
the aligned (u/u) case indeed has a higher associated
power.

In all cases, a significant non-collimated
emission is induced (illustrated in the top right plot of Fig.~\ref{fig:binary_field}) during the merger phase. Clearly,
the simple minded picture of a jet produced by the superposition of the orbital
and spinning effect can not fully capture the complete behavior 
at the merger epoch, although it serves to understand
the main qualitative features and provides a means to estimate the power of the electromagnetic
emission.

\begin{figure}
$\begin{array}{c@{\hspace{0.1in}}c@{\hspace{0.1in}}c}
\includegraphics[width=1.7in]{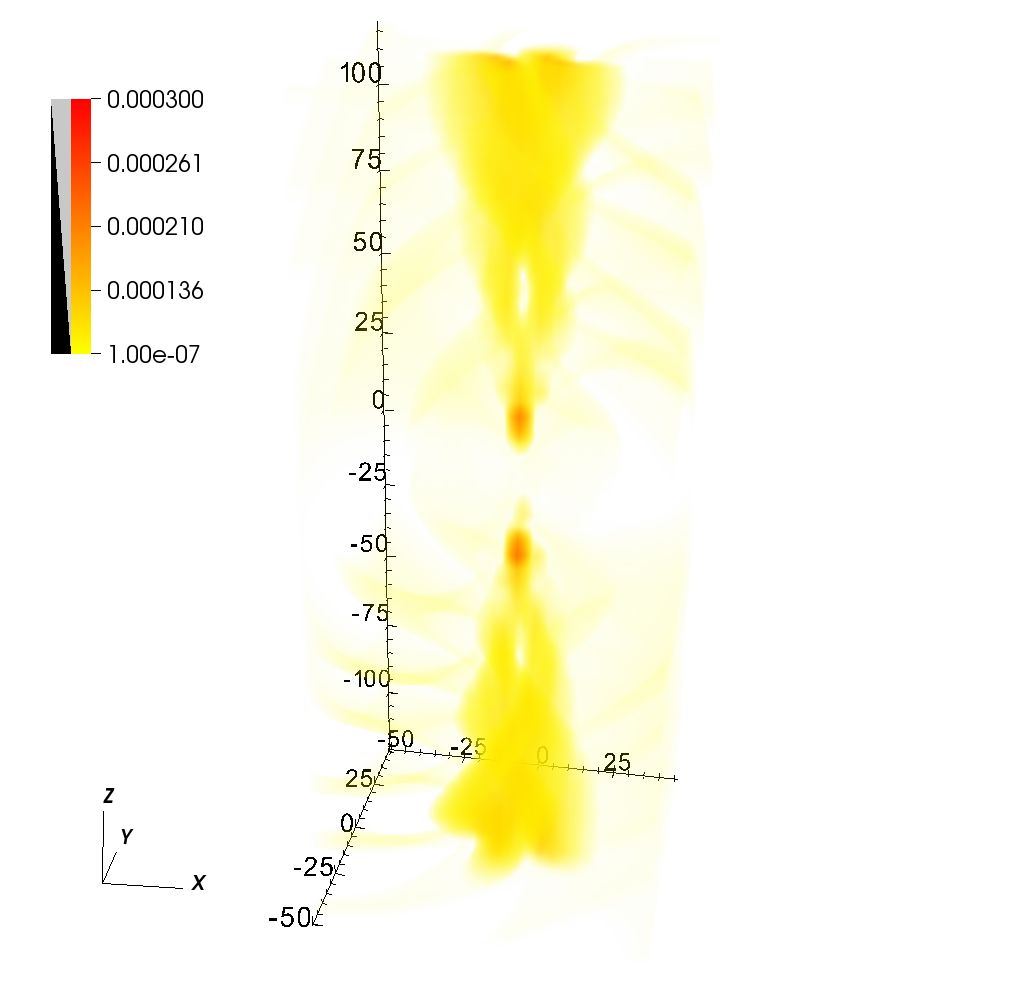} &
\includegraphics[width=1.7in]{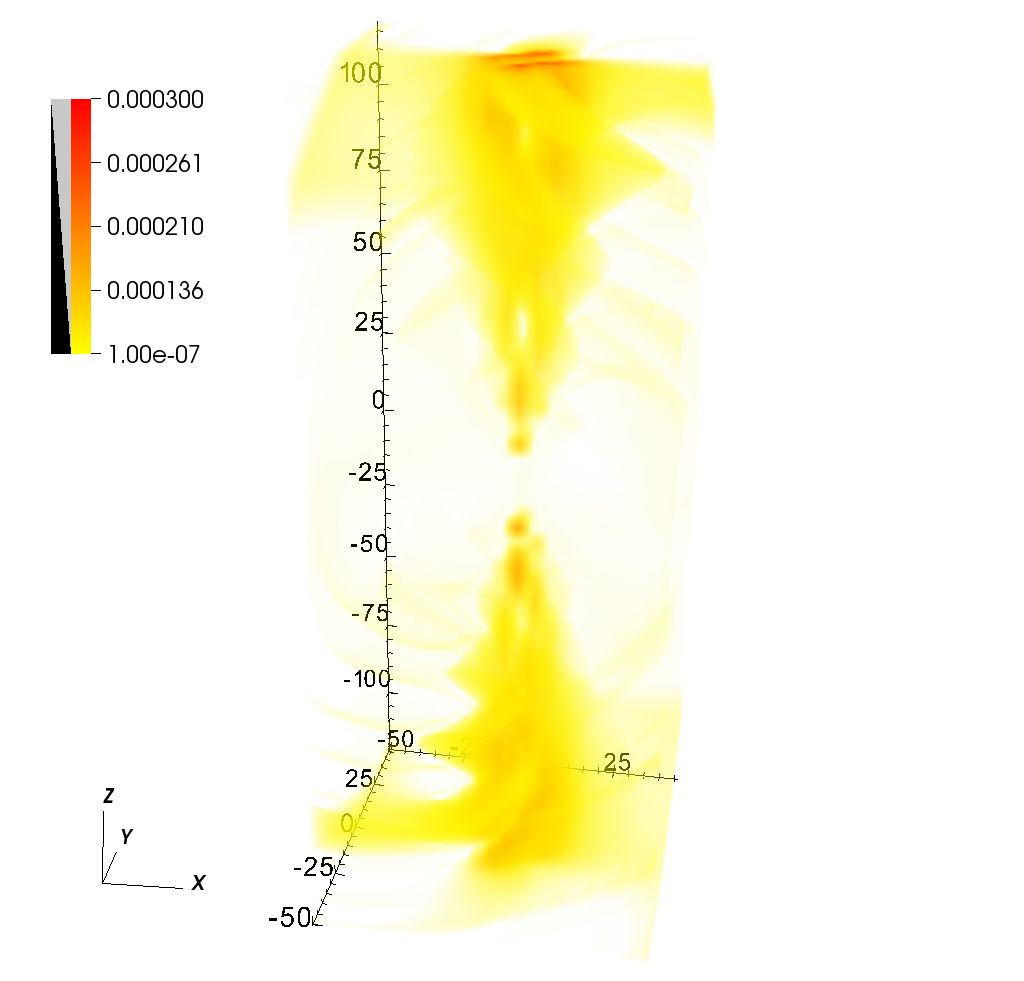} \\
\includegraphics[width=1.7in]{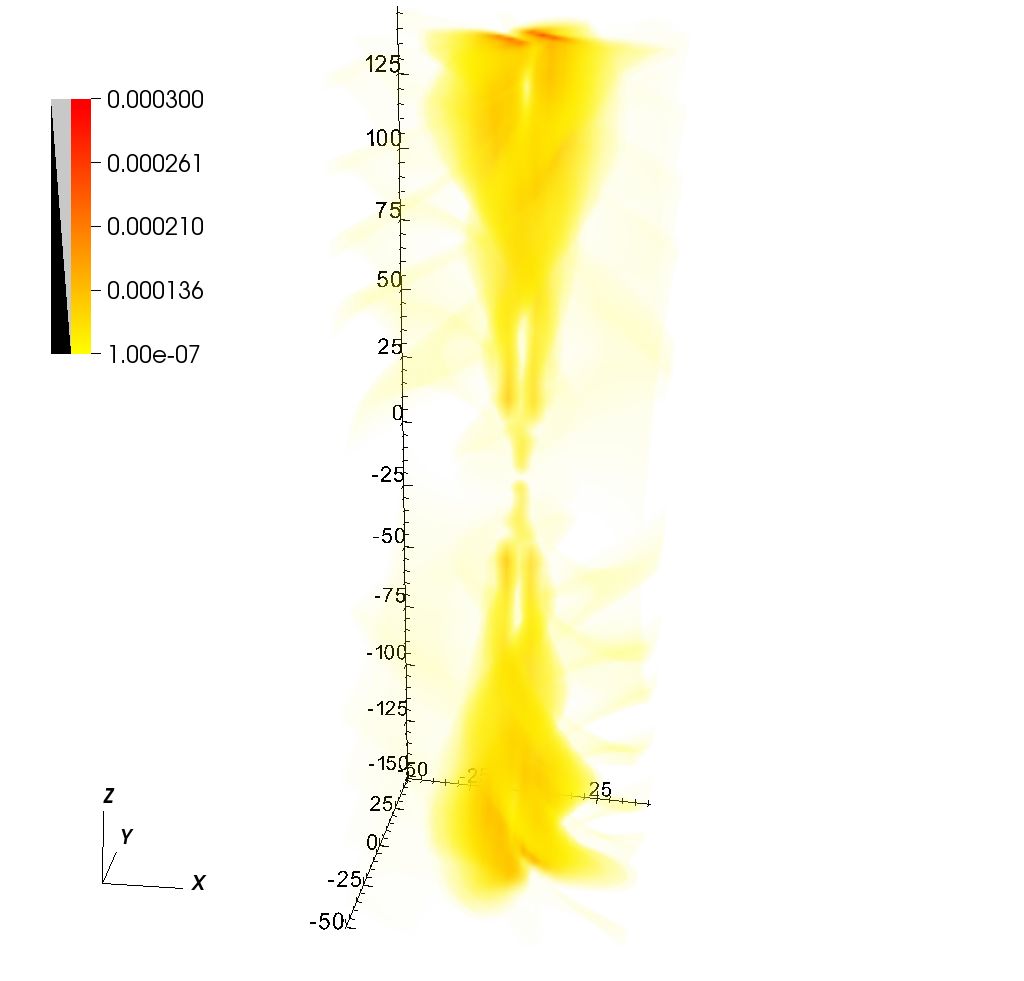} &
\includegraphics[width=1.7in]{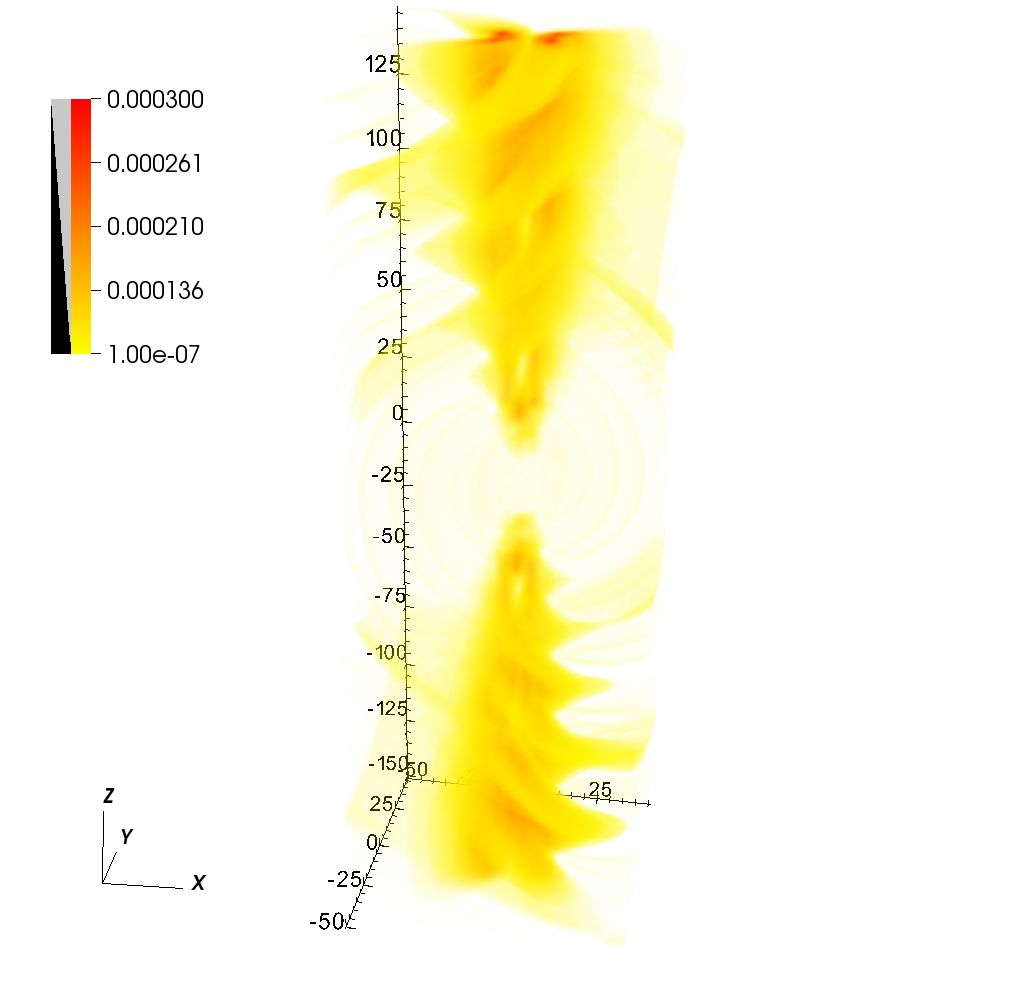} &
\end{array}$
\begin{caption}
{Electromagnetic flux corresponding to the 0/0 (top row) and u/d (bottom row) cases 
at times $(-16.7, -2.5)$ ${\rm hrs} \, M_8$ and $(-10.5, 3.6)$ ${\rm hrs} \, M_8$ (with respect to
the merger time as marked by the peak in strength of gravitational wave emission) 
respectively. The orbiting stage leaves its
imprint as twisted tubes.}
\end{caption}
\end{figure}

\begin{figure}
$\begin{array}{c@{\hspace{0.1in}}c@{\hspace{0.1in}}c}
\includegraphics[width=1.7in]{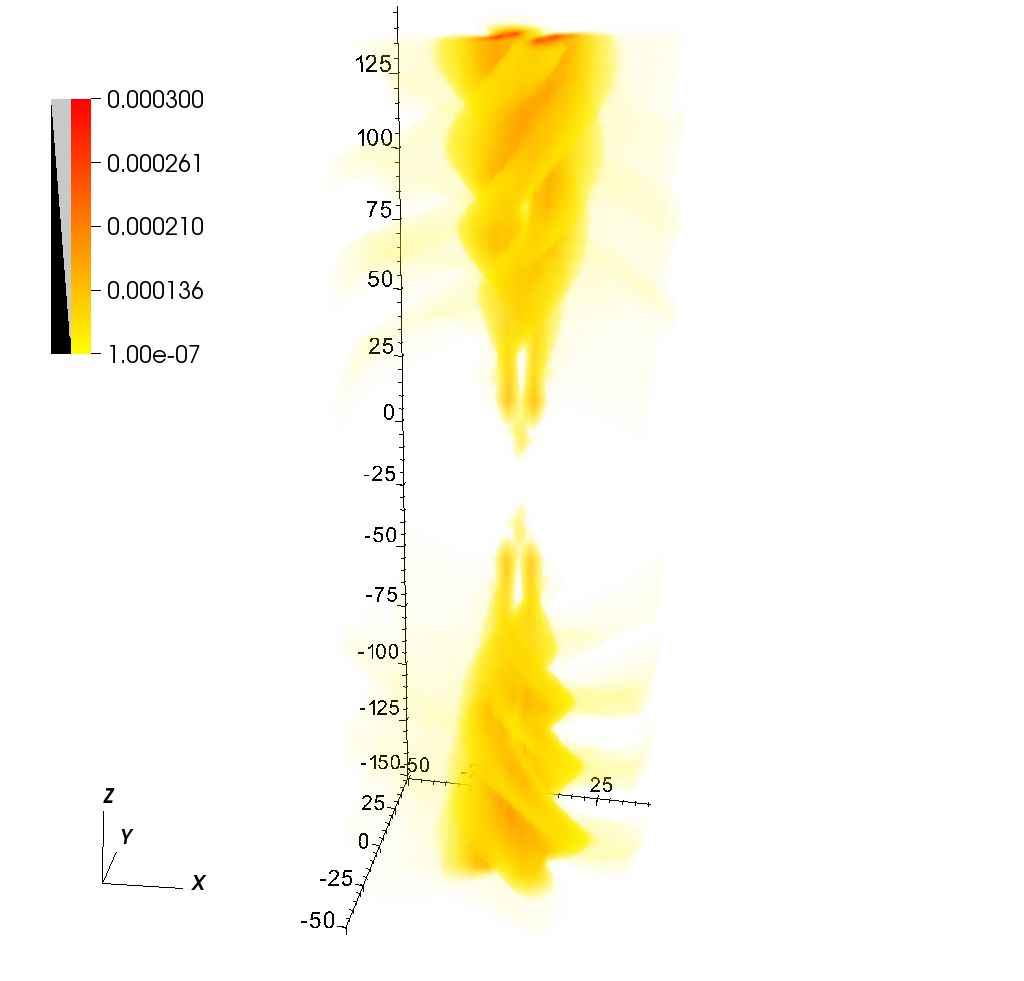} &
\includegraphics[width=1.7in]{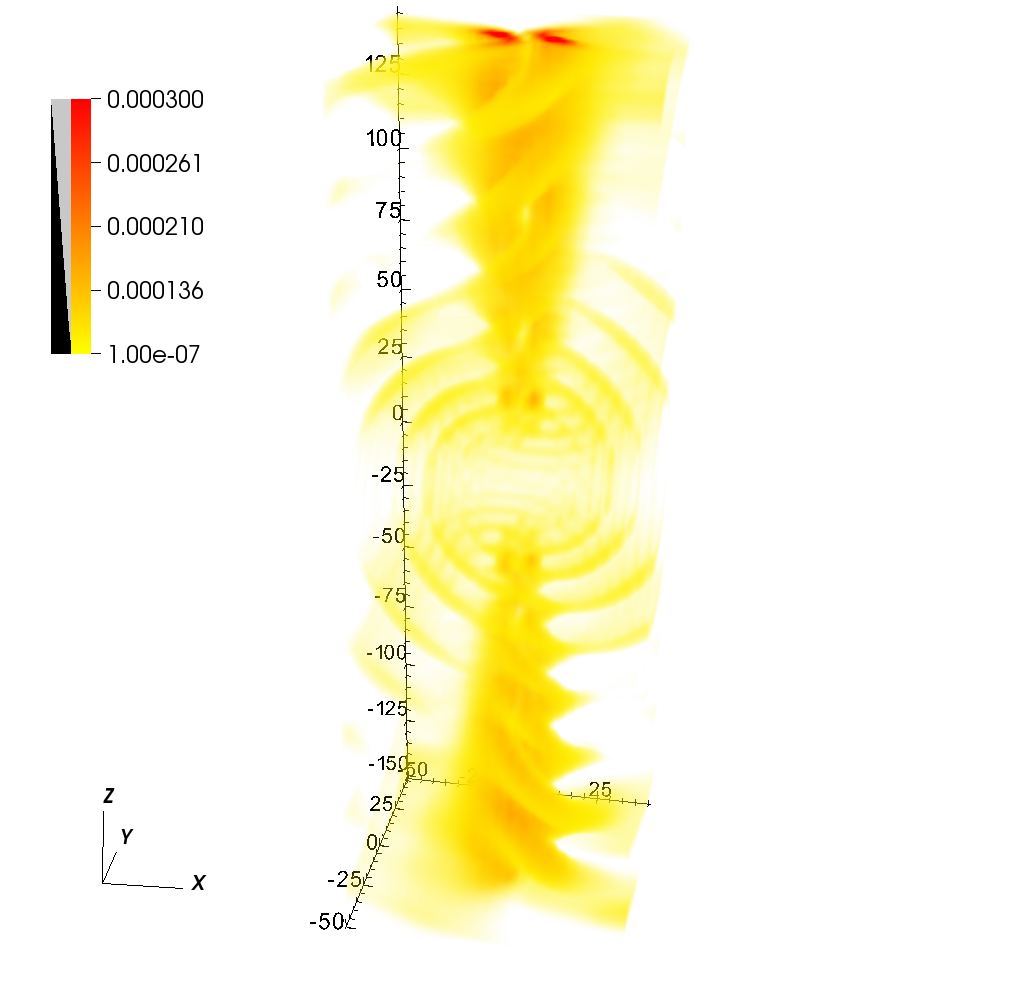} \\
\includegraphics[width=1.7in]{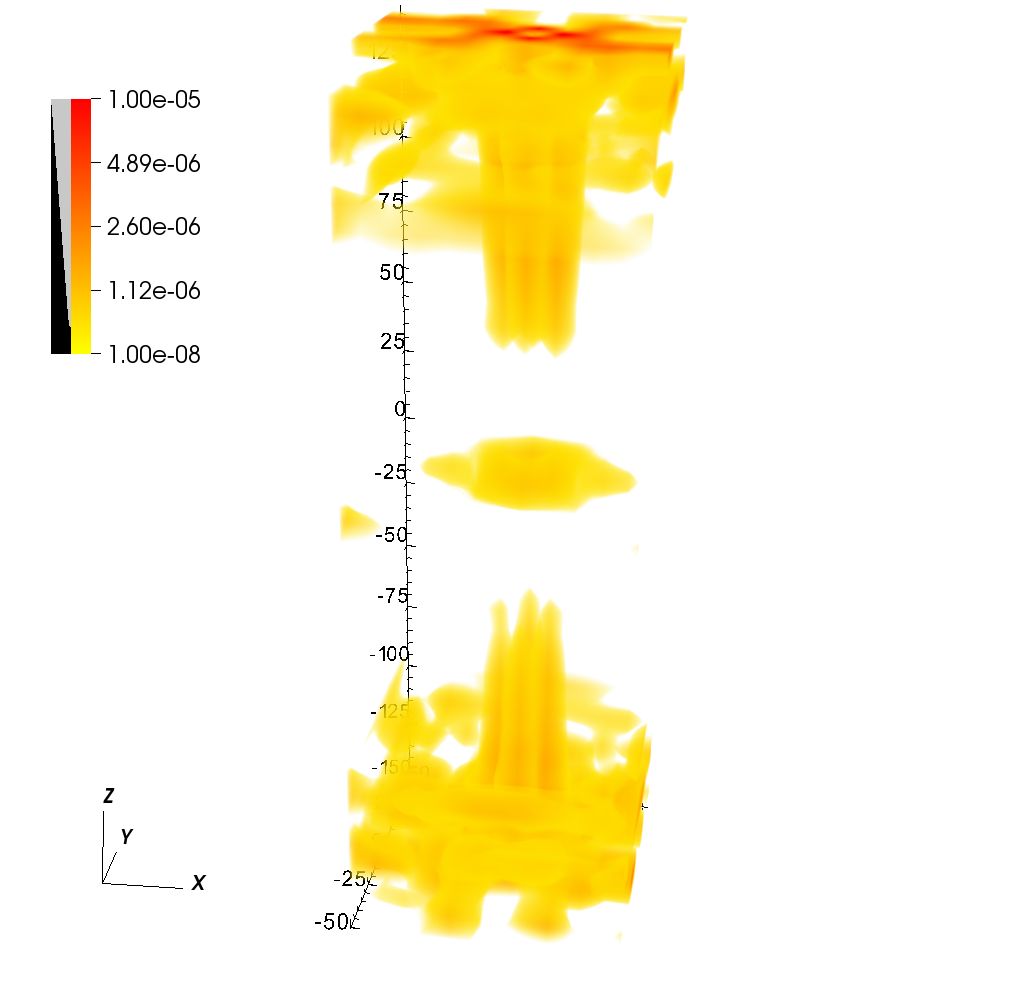} &
\includegraphics[width=1.7in]{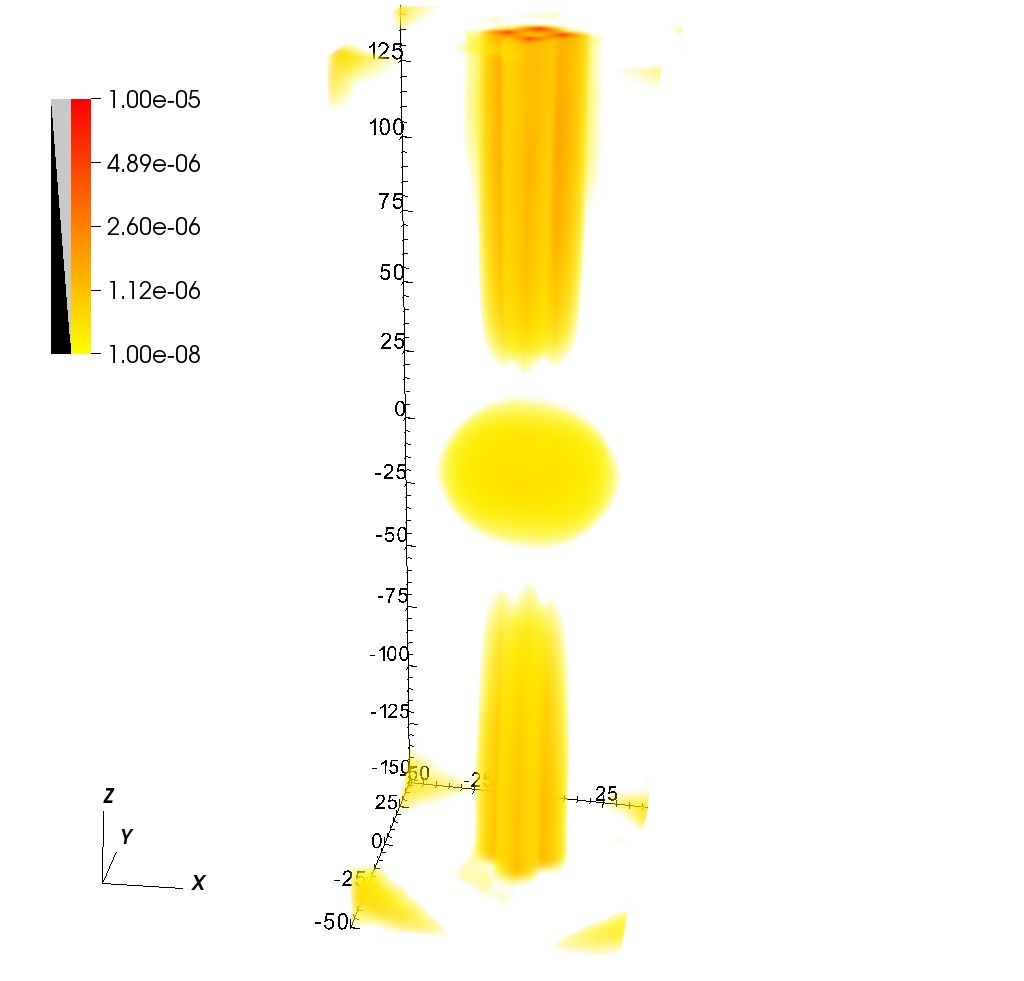}
\end{array}$
\begin{caption}
 {Electromagnetic flux corresponding to the u/u case at times $(-2.5, 11.6, 40, 68)$ ${\rm hrs} \, M_8$
(with respect to
the merger time as marked by the peak in strength of gravitational wave emission).
The orbiting behavior leaves a clear imprint in the resulting
flux of energy and, as the merger proceeds emission along all directions
is evident. At late times, the system settles to a single jet as 
dictated by the BZ mechanism.}
\end{caption}
\end{figure}

%
%
\section{Final Comments}
\label{sec:conclusions}
We have studied the impact of black hole motion through a plasma and indicated
how the interaction can induce powerful electromagnetic emissions even for
non-spinning black holes.
%
Despite having examined a very small subset of the binary black hole
parameter space, the results presented both here and
in~\cite{sciencejets,prdjets} suggest a wide applicability to general
black hole binaries. Moreover, as the plasma generally has
a negligible effect on the dynamics of the black holes, then one needs
only to know
the dynamics of the black holes, say by numerical solution or other approximate
methods, in order to estimate the expected electromagnetic luminosity.
A recent example is the work of~\cite{McWilliams:2010iq} which uses the expected
 $\simeq B^2 v^2$
scaling for non-spinning black hole binaries  coupled with the known distance dependence
in time to obtain excellent agreement with the obtained luminosity from such a system.

Naturally, spinning black holes produce stronger jets, and 
these jets, as shown earlier~\cite{prdjets}, will be aligned 
with the asymptotic 
magnetic field direction\footnote{The magnitude of the BZ-associated emission 
diminishes for non-aligned cases but
it is nevertheless significant: the orthogonal case is only 
half as powerful as the aligned case.}.
 Thus, the resulting electromagnetic luminosity can be estimated
to be $L^{EM}_{col} \simeq L_{\rm spin} + L_{\rm speed}$, 
which can be significant and have
associated time-variabilities tied to the dynamical behavior of the system. 
 In particular, the luminosity tied to the motion can result significantly
higher than that tied to the spin.
In addition, eccentric orbits
and spin-orbit interactions driving orbital plane precessions 
can induce important variabilities
that can aid in the detection of these systems. Furthermore,
a significant pulse of nearly isotropic radiation is emitted during merger, 
thereby allowing
observations of the system along directions not aligned with the jet.

Consequently, binary black hole interactions with surrounding plasmas can yield powerful electromagnetic outputs
and allow for observing these systems through both
gravitational and electromagnetic radiation.
Gravitational waves from these systems corresponding to the last year
before the merger could be observed to large distances with LISA (up to
redshifts of 5-10~\cite{Vecchio:2003tn}) for
masses $\simeq 10^{4-7} M_{\odot}$ or earlier in the orbiting phase (and possibly
through merger) via Pulsar Timing Array observations~\cite{Demorest:2009ex}, targeting binaries
with masses in $M \simeq 10^{7 - 10} M_{\odot}$. As we have indicated here, both scenarios can have 
strong associated electromagnetic emissions. Our inferred luminosities of several $10^{43}
(M_8 B_4)^2 $ ergs/s
corresponds to an isotropic bolometric
flux of $F_x \simeq 10^{-15}$ erg $(M_8 B_4)^2 $ /(${\rm cm}^2$ s) that could be detected to redshifts of
$z \approx 1$  and even further depending on anisotropies, depending on the efficiency
of processes taping this available energy and producing observable signals.

%
%
\section*{Acknowledgments}
It is a pleasure to thank  J. Aarons, P. Chang,
B. MacNamara, K. Menou, E. Quataert and C. Thompson as well as
our long time collaborators Matthew Anderson, Miguel Megevand and Oscar Reula
for useful discussions and comments. We acknowledge
support from NSF grants PHY-0803629 to Louisiana State University,
PHY-0969811 to Brigham Young University,
PHY-0969827 to Long Island University, 
as well as NSERC through a Discovery Grant.
Research at Perimeter Institute is
supported through Industry Canada and by the Province of Ontario
through the Ministry of Research \& Innovation.  Computations were
performed at LONI, Teragrid and Scinet.

%
%
\bibliographystyle{apsrev}

\begin{thebibliography}{41}
\expandafter\ifx\csname natexlab\endcsname\relax\def\natexlab#1{#1}\fi
\expandafter\ifx\csname bibnamefont\endcsname\relax
  \def\bibnamefont#1{#1}\fi
\expandafter\ifx\csname bibfnamefont\endcsname\relax
  \def\bibfnamefont#1{#1}\fi
\expandafter\ifx\csname citenamefont\endcsname\relax
  \def\citenamefont#1{#1}\fi
\expandafter\ifx\csname url\endcsname\relax
  \def\url#1{\texttt{#1}}\fi
\expandafter\ifx\csname urlprefix\endcsname\relax\def\urlprefix{URL }\fi
\providecommand{\bibinfo}[2]{#2}
\providecommand{\eprint}[2][]{\url{#2}}

\bibitem[{\citenamefont{Hamuy}(2003)}]{Hamuy:2003xv}
\bibinfo{author}{\bibfnamefont{M.}~\bibnamefont{Hamuy}} (\bibinfo{year}{2003}),
  \eprint{astro-ph/0301006}.

\bibitem[{\citenamefont{Woosley and Bloom}(2006)}]{Woosley:2006fn}
\bibinfo{author}{\bibfnamefont{S.~E.} \bibnamefont{Woosley}} \bibnamefont{and}
  \bibinfo{author}{\bibfnamefont{J.~S.} \bibnamefont{Bloom}},
  \bibinfo{journal}{Ann. Rev. Astron. Astrophys.}
  \textbf{\bibinfo{volume}{44}}, \bibinfo{pages}{507} (\bibinfo{year}{2006}),
  \eprint{astro-ph/0609142}.

\bibitem[{\citenamefont{{McNamara} et~al.}(2001)\citenamefont{{McNamara},
  {Wise}, {Nulsen}, {David}, {Carilli}, {Sarazin}, {O'Dea}, {Houck}, {Donahue},
  {Baum} et~al.}}]{2001ApJ...562L.149M}
\bibinfo{author}{\bibfnamefont{B.~R.} \bibnamefont{{McNamara}}},
  \bibinfo{author}{\bibfnamefont{M.~W.} \bibnamefont{{Wise}}},
  \bibinfo{author}{\bibfnamefont{P.~E.~J.} \bibnamefont{{Nulsen}}},
  \bibinfo{author}{\bibfnamefont{L.~P.} \bibnamefont{{David}}},
  \bibinfo{author}{\bibfnamefont{C.~L.} \bibnamefont{{Carilli}}},
  \bibinfo{author}{\bibfnamefont{C.~L.} \bibnamefont{{Sarazin}}},
  \bibinfo{author}{\bibfnamefont{C.~P.} \bibnamefont{{O'Dea}}},
  \bibinfo{author}{\bibfnamefont{J.}~\bibnamefont{{Houck}}},
  \bibinfo{author}{\bibfnamefont{M.}~\bibnamefont{{Donahue}}},
  \bibinfo{author}{\bibfnamefont{S.}~\bibnamefont{{Baum}}},
  \bibnamefont{et~al.}, \bibinfo{journal}{\apjl}
  \textbf{\bibinfo{volume}{562}}, \bibinfo{pages}{L149} (\bibinfo{year}{2001}),
  \eprint{arXiv:astro-ph/0110554}.

\bibitem[{\citenamefont{Fender et~al.}(2010)\citenamefont{Fender, Gallo, and
  Russell}}]{Fender:2010tk}
\bibinfo{author}{\bibfnamefont{R.}~\bibnamefont{Fender}},
  \bibinfo{author}{\bibfnamefont{E.}~\bibnamefont{Gallo}}, \bibnamefont{and}
  \bibinfo{author}{\bibfnamefont{D.}~\bibnamefont{Russell}}
  (\bibinfo{year}{2010}), \eprint{1003.5516}.

\bibitem[{\citenamefont{Penrose}(1969)}]{Penrose:1969pc}
\bibinfo{author}{\bibfnamefont{R.}~\bibnamefont{Penrose}},
  \bibinfo{journal}{Riv. Nuovo Cim.} \textbf{\bibinfo{volume}{1}},
  \bibinfo{pages}{252} (\bibinfo{year}{1969}).

\bibitem[{\citenamefont{Blandford and Znajek}(1977)}]{Blandford:1977ds}
\bibinfo{author}{\bibfnamefont{R.~D.} \bibnamefont{Blandford}}
  \bibnamefont{and} \bibinfo{author}{\bibfnamefont{R.~L.}
  \bibnamefont{Znajek}}, \bibinfo{journal}{Mon. Not. Roy. Astron. Soc.}
  \textbf{\bibinfo{volume}{179}}, \bibinfo{pages}{433} (\bibinfo{year}{1977}).

\bibitem[{\citenamefont{Lee et~al.}(2000)\citenamefont{Lee, Wijers, and
  Brown}}]{2000PhR...325...83L}
\bibinfo{author}{\bibfnamefont{H.~K.} \bibnamefont{Lee}},
  \bibinfo{author}{\bibfnamefont{R.~A.~M.~J.} \bibnamefont{Wijers}},
  \bibnamefont{and} \bibinfo{author}{\bibfnamefont{G.~E.} \bibnamefont{Brown}},
  \bibinfo{journal}{\physrep} \textbf{\bibinfo{volume}{325}},
  \bibinfo{pages}{83} (\bibinfo{year}{2000}), \eprint{arXiv:astro-ph/9906213}.

\bibitem[{\citenamefont{{Blandford}}(2002)}]{2002luml.conf..381B}
\bibinfo{author}{\bibfnamefont{R.~D.} \bibnamefont{{Blandford}}}, in
  \emph{\bibinfo{booktitle}{Lighthouses of the Universe: The Most Luminous
  Celestial Objects and Their Use for Cosmology}}, edited by
  \bibinfo{editor}{\bibnamefont{{M.~Gilfanov, R.~Sunyeav, \& E.~Churazov}}}
  (\bibinfo{year}{2002}), pp. \bibinfo{pages}{381--+}.

\bibitem[{\citenamefont{{Punsly}}(2008)}]{2008ASSL..355.....P}
\bibinfo{editor}{\bibfnamefont{B.}~\bibnamefont{{Punsly}}}, ed.,
  \emph{\bibinfo{title}{{Black Hole Gravitohydromagnetics}}}, vol.
  \bibinfo{volume}{355} of \emph{\bibinfo{series}{Astrophysics and Space
  Science Library}} (\bibinfo{year}{2008}).

\bibitem[{\citenamefont{Palenzuela
  et~al.}(2010{\natexlab{a}})\citenamefont{Palenzuela, Lehner, and
  Liebling}}]{sciencejets}
\bibinfo{author}{\bibfnamefont{C.}~\bibnamefont{Palenzuela}},
  \bibinfo{author}{\bibfnamefont{L.}~\bibnamefont{Lehner}}, \bibnamefont{and}
  \bibinfo{author}{\bibfnamefont{S.~L.} \bibnamefont{Liebling}},
  \bibinfo{journal}{Science} \textbf{\bibinfo{volume}{329}},
  \bibinfo{pages}{927} (\bibinfo{year}{2010}{\natexlab{a}}),
  \eprint{1005.1067}.

\bibitem[{\citenamefont{Palenzuela
  et~al.}(2010{\natexlab{b}})\citenamefont{Palenzuela, Garrett, Lehner, and
  Liebling}}]{prdjets}
\bibinfo{author}{\bibfnamefont{C.}~\bibnamefont{Palenzuela}},
  \bibinfo{author}{\bibfnamefont{T.}~\bibnamefont{Garrett}},
  \bibinfo{author}{\bibfnamefont{L.}~\bibnamefont{Lehner}}, \bibnamefont{and}
  \bibinfo{author}{\bibfnamefont{S.~L.} \bibnamefont{Liebling}},
  \bibinfo{journal}{Phys. Rev.} \textbf{\bibinfo{volume}{D82}},
  \bibinfo{pages}{044045} (\bibinfo{year}{2010}{\natexlab{b}}),
  \eprint{1007.1198}.

\bibitem[{\citenamefont{{Begelman} et~al.}(1980)\citenamefont{{Begelman},
  {Blandford}, and {Rees}}}]{1980Natur.287..307B}
\bibinfo{author}{\bibfnamefont{M.~C.} \bibnamefont{{Begelman}}},
  \bibinfo{author}{\bibfnamefont{R.~D.} \bibnamefont{{Blandford}}},
  \bibnamefont{and} \bibinfo{author}{\bibfnamefont{M.~J.}
  \bibnamefont{{Rees}}}, \bibinfo{journal}{\nat}
  \textbf{\bibinfo{volume}{287}}, \bibinfo{pages}{307} (\bibinfo{year}{1980}).

\bibitem[{\citenamefont{Milosavljevic and
  Phinney}(2005)}]{Milosavljevic:2004cg}
\bibinfo{author}{\bibfnamefont{M.}~\bibnamefont{Milosavljevic}}
  \bibnamefont{and} \bibinfo{author}{\bibfnamefont{E.~S.}
  \bibnamefont{Phinney}}, \bibinfo{journal}{Astrophys. J.}
  \textbf{\bibinfo{volume}{622}}, \bibinfo{pages}{L93} (\bibinfo{year}{2005}).

\bibitem[{\citenamefont{Comerford et~al.}(2009)}]{Comerford:2008gm}
\bibinfo{author}{\bibfnamefont{J.~M.} \bibnamefont{Comerford}}
  \bibnamefont{et~al.}, \bibinfo{journal}{Astrophys. J.}
  \textbf{\bibinfo{volume}{698}}, \bibinfo{pages}{956} (\bibinfo{year}{2009}),
  \eprint{0810.3235}.

\bibitem[{\citenamefont{{Schutz}}(2002)}]{2002luml.conf..207S}
\bibinfo{author}{\bibfnamefont{B.~F.} \bibnamefont{{Schutz}}}, in
  \emph{\bibinfo{booktitle}{Lighthouses of the Universe: The Most Luminous
  Celestial Objects and Their Use for Cosmology}}, edited by
  \bibinfo{editor}{\bibnamefont{{M.~Gilfanov, R.~Sunyeav, \& E.~Churazov}}}
  (\bibinfo{year}{2002}), pp. \bibinfo{pages}{207--+}.

\bibitem[{\citenamefont{{Sylvestre}}(2003)}]{2003ApJ...591.1152S}
\bibinfo{author}{\bibfnamefont{J.}~\bibnamefont{{Sylvestre}}},
  \bibinfo{journal}{\apj} \textbf{\bibinfo{volume}{591}}, \bibinfo{pages}{1152}
  (\bibinfo{year}{2003}), \eprint{arXiv:astro-ph/0303512}.

\bibitem[{\citenamefont{{Bloom} et~al.}(2009)\citenamefont{{Bloom}, {Holz},
  {Hughes}, {Menou}, {Adams}, {Anderson}, {Becker}, {Bower}, {Brandt}, {Cobb}
  et~al.}}]{2009arXiv0902.1527B}
\bibinfo{author}{\bibfnamefont{J.~S.} \bibnamefont{{Bloom}}},
  \bibinfo{author}{\bibfnamefont{D.~E.} \bibnamefont{{Holz}}},
  \bibinfo{author}{\bibfnamefont{S.~A.} \bibnamefont{{Hughes}}},
  \bibinfo{author}{\bibfnamefont{K.}~\bibnamefont{{Menou}}},
  \bibinfo{author}{\bibfnamefont{A.}~\bibnamefont{{Adams}}},
  \bibinfo{author}{\bibfnamefont{S.~F.} \bibnamefont{{Anderson}}},
  \bibinfo{author}{\bibfnamefont{A.}~\bibnamefont{{Becker}}},
  \bibinfo{author}{\bibfnamefont{G.~C.} \bibnamefont{{Bower}}},
  \bibinfo{author}{\bibfnamefont{N.}~\bibnamefont{{Brandt}}},
  \bibinfo{author}{\bibfnamefont{B.}~\bibnamefont{{Cobb}}},
  \bibnamefont{et~al.}, \bibinfo{journal}{ArXiv e-prints}
  (\bibinfo{year}{2009}), \eprint{0902.1527}.

\bibitem[{\citenamefont{{Phinney}}(2009)}]{2009astro2010S.235P}
\bibinfo{author}{\bibfnamefont{E.~S.} \bibnamefont{{Phinney}}}, in
  \emph{\bibinfo{booktitle}{astro2010: The Astronomy and Astrophysics Decadal
  Survey}} (\bibinfo{year}{2009}), vol. \bibinfo{volume}{2010} of
  \emph{\bibinfo{series}{Astronomy}}, pp. \bibinfo{pages}{235--+}.

\bibitem[{\citenamefont{{Goldreich} and {Julian}}(1969)}]{1969ApJ...157..869G}
\bibinfo{author}{\bibfnamefont{P.}~\bibnamefont{{Goldreich}}} \bibnamefont{and}
  \bibinfo{author}{\bibfnamefont{W.~H.} \bibnamefont{{Julian}}},
  \bibinfo{journal}{\apj} \textbf{\bibinfo{volume}{157}}, \bibinfo{pages}{869}
  (\bibinfo{year}{1969}).

\bibitem[{\citenamefont{Baumgarte and Shapiro}(1999)}]{Baumgarte:1998te}
\bibinfo{author}{\bibfnamefont{T.~W.} \bibnamefont{Baumgarte}}
  \bibnamefont{and} \bibinfo{author}{\bibfnamefont{S.~L.}
  \bibnamefont{Shapiro}}, \bibinfo{journal}{Phys. Rev.}
  \textbf{\bibinfo{volume}{D59}}, \bibinfo{pages}{024007}
  (\bibinfo{year}{1999}), \eprint{gr-qc/9810065}.

\bibitem[{\citenamefont{Shibata and Nakamura}(1995)}]{Shibata:1995we}
\bibinfo{author}{\bibfnamefont{M.}~\bibnamefont{Shibata}} \bibnamefont{and}
  \bibinfo{author}{\bibfnamefont{T.}~\bibnamefont{Nakamura}},
  \bibinfo{journal}{Phys. Rev.} \textbf{\bibinfo{volume}{D52}},
  \bibinfo{pages}{5428} (\bibinfo{year}{1995}).

\bibitem[{had()}]{had_webpage}
\bibinfo{note}{Http://www.had.liu.edu/}.

\bibitem[{\citenamefont{Liebling}(2002)}]{Liebling}
\bibinfo{author}{\bibfnamefont{S.~L.} \bibnamefont{Liebling}},
  \bibinfo{journal}{Phys. Rev.} \textbf{\bibinfo{volume}{D66}},
  \bibinfo{pages}{041703} (\bibinfo{year}{2002}).

\bibitem[{\citenamefont{Lehner et~al.}(2006)\citenamefont{Lehner, Liebling, and
  Reula}}]{Lehner:2005vc}
\bibinfo{author}{\bibfnamefont{L.}~\bibnamefont{Lehner}},
  \bibinfo{author}{\bibfnamefont{S.~L.} \bibnamefont{Liebling}},
  \bibnamefont{and} \bibinfo{author}{\bibfnamefont{O.}~\bibnamefont{Reula}},
  \bibinfo{journal}{Class. Quant. Grav.} \textbf{\bibinfo{volume}{23}},
  \bibinfo{pages}{S421} (\bibinfo{year}{2006}).

\bibitem[{\citenamefont{Pretorius}(2002)}]{pretorius}
\bibinfo{author}{\bibfnamefont{F.}~\bibnamefont{Pretorius}}, Ph.D. thesis,
  \bibinfo{school}{The University of British Columbia} (\bibinfo{year}{2002}).

\bibitem[{\citenamefont{Anderson et~al.}(2008)}]{binaryns}
\bibinfo{author}{\bibfnamefont{M.}~\bibnamefont{Anderson}}
  \bibnamefont{et~al.}, \bibinfo{journal}{Phys. Rev.}
  \textbf{\bibinfo{volume}{D77}}, \bibinfo{pages}{024006}
  (\bibinfo{year}{2008}).

\bibitem[{\citenamefont{Newman and Penrose}(1962)}]{Newman:1961qr}
\bibinfo{author}{\bibfnamefont{E.}~\bibnamefont{Newman}} \bibnamefont{and}
  \bibinfo{author}{\bibfnamefont{R.}~\bibnamefont{Penrose}},
  \bibinfo{journal}{J. Math. Phys.} \textbf{\bibinfo{volume}{3}},
  \bibinfo{pages}{566} (\bibinfo{year}{1962}).

\bibitem[{\citenamefont{Lehner and Moreschi}(2007)}]{Lehner:2007ip}
\bibinfo{author}{\bibfnamefont{L.}~\bibnamefont{Lehner}} \bibnamefont{and}
  \bibinfo{author}{\bibfnamefont{O.~M.} \bibnamefont{Moreschi}},
  \bibinfo{journal}{Phys. Rev.} \textbf{\bibinfo{volume}{D76}},
  \bibinfo{pages}{124040} (\bibinfo{year}{2007}), \eprint{0706.1319}.

\bibitem[{\citenamefont{Massi and Kaufman}(2008)}]{2008A&A...477....1M}
\bibinfo{author}{\bibfnamefont{M.}~\bibnamefont{Massi}} \bibnamefont{and}
  \bibinfo{author}{\bibfnamefont{M.}~\bibnamefont{Kaufman}},
  \bibinfo{journal}{Astronomy and Astrophysics} \textbf{\bibinfo{volume}{477}},
  \bibinfo{pages}{1} (\bibinfo{year}{2008}).

\bibitem[{\citenamefont{{Field} and {Rogers}}(1993)}]{1993ApJ...403...94F}
\bibinfo{author}{\bibfnamefont{G.~B.} \bibnamefont{{Field}}} \bibnamefont{and}
  \bibinfo{author}{\bibfnamefont{R.~D.} \bibnamefont{{Rogers}}},
  \bibinfo{journal}{\apj} \textbf{\bibinfo{volume}{403}}, \bibinfo{pages}{94}
  (\bibinfo{year}{1993}).

\bibitem[{\citenamefont{{Dermer} et~al.}(2008)\citenamefont{{Dermer}, {Finke},
  and {Menon}}}]{2008arXiv0810.1055D}
\bibinfo{author}{\bibfnamefont{C.~D.} \bibnamefont{{Dermer}}},
  \bibinfo{author}{\bibfnamefont{J.~D.} \bibnamefont{{Finke}}},
  \bibnamefont{and} \bibinfo{author}{\bibfnamefont{G.}~\bibnamefont{{Menon}}},
  \bibinfo{journal}{ArXiv e-prints}  (\bibinfo{year}{2008}),
  \eprint{0810.1055}.

\bibitem[{\citenamefont{{Tchekhovskoy}
  et~al.}(2010)\citenamefont{{Tchekhovskoy}, {Narayan}, and
  {McKinney}}}]{2010ApJ...711...50T}
\bibinfo{author}{\bibfnamefont{A.}~\bibnamefont{{Tchekhovskoy}}},
  \bibinfo{author}{\bibfnamefont{R.}~\bibnamefont{{Narayan}}},
  \bibnamefont{and} \bibinfo{author}{\bibfnamefont{J.~C.}
  \bibnamefont{{McKinney}}}, \bibinfo{journal}{\apj}
  \textbf{\bibinfo{volume}{711}}, \bibinfo{pages}{50} (\bibinfo{year}{2010}),
  \eprint{0911.2228}.

\bibitem[{\citenamefont{Palenzuela et~al.}(2009)\citenamefont{Palenzuela,
  Anderson, Lehner, Liebling, and Neilsen}}]{Palenzuela:2009yr}
\bibinfo{author}{\bibfnamefont{C.}~\bibnamefont{Palenzuela}},
  \bibinfo{author}{\bibfnamefont{M.}~\bibnamefont{Anderson}},
  \bibinfo{author}{\bibfnamefont{L.}~\bibnamefont{Lehner}},
  \bibinfo{author}{\bibfnamefont{S.~L.} \bibnamefont{Liebling}},
  \bibnamefont{and} \bibinfo{author}{\bibfnamefont{D.}~\bibnamefont{Neilsen}},
  \bibinfo{journal}{Phys. Rev. Lett.} \textbf{\bibinfo{volume}{103}},
  \bibinfo{pages}{081101} (\bibinfo{year}{2009}), \eprint{0905.1121}.

\bibitem[{\citenamefont{Palenzuela
  et~al.}(2010{\natexlab{c}})\citenamefont{Palenzuela, Lehner, and
  Yoshida}}]{Palenzuela:2009hx}
\bibinfo{author}{\bibfnamefont{C.}~\bibnamefont{Palenzuela}},
  \bibinfo{author}{\bibfnamefont{L.}~\bibnamefont{Lehner}}, \bibnamefont{and}
  \bibinfo{author}{\bibfnamefont{S.}~\bibnamefont{Yoshida}},
  \bibinfo{journal}{Phys. Rev.} \textbf{\bibinfo{volume}{D81}},
  \bibinfo{pages}{084007} (\bibinfo{year}{2010}{\natexlab{c}}),
  \eprint{0911.3889}.

\bibitem[{\citenamefont{{Thorne} et~al.}(1986)\citenamefont{{Thorne}, {Price},
  and {MacDonald}}}]{1986bhmp.book.....T}
\bibinfo{author}{\bibfnamefont{K.~S.} \bibnamefont{{Thorne}}},
  \bibinfo{author}{\bibfnamefont{R.~H.} \bibnamefont{{Price}}},
  \bibnamefont{and} \bibinfo{author}{\bibfnamefont{D.~A.}
  \bibnamefont{{MacDonald}}}, \emph{\bibinfo{title}{{Black holes: The membrane
  paradigm}}} (\bibinfo{year}{1986}).

\bibitem[{\citenamefont{Drell et~al.}(1965)\citenamefont{Drell, Foley, and
  Ruderman}}]{1965JGR....70.3131D}
\bibinfo{author}{\bibfnamefont{S.~D.} \bibnamefont{Drell}},
  \bibinfo{author}{\bibfnamefont{H.~M.} \bibnamefont{Foley}}, \bibnamefont{and}
  \bibinfo{author}{\bibfnamefont{M.~A.} \bibnamefont{Ruderman}},
  \bibinfo{journal}{Journal of Geophysical Research}
  \textbf{\bibinfo{volume}{70}}, \bibinfo{pages}{3131} (\bibinfo{year}{1965}).

\bibitem[{\citenamefont{Hannam et~al.}(2009)}]{Hannam:2009hh}
\bibinfo{author}{\bibfnamefont{M.}~\bibnamefont{Hannam}} \bibnamefont{et~al.},
  \bibinfo{journal}{Phys. Rev.} \textbf{\bibinfo{volume}{D79}},
  \bibinfo{pages}{084025} (\bibinfo{year}{2009}), \eprint{0901.2437}.

\bibitem[{\citenamefont{Buonanno et~al.}(2008)\citenamefont{Buonanno, Kidder,
  and Lehner}}]{Buonanno:2007sv}
\bibinfo{author}{\bibfnamefont{A.}~\bibnamefont{Buonanno}},
  \bibinfo{author}{\bibfnamefont{L.~E.} \bibnamefont{Kidder}},
  \bibnamefont{and} \bibinfo{author}{\bibfnamefont{L.}~\bibnamefont{Lehner}},
  \bibinfo{journal}{Phys. Rev.} \textbf{\bibinfo{volume}{D77}},
  \bibinfo{pages}{026004} (\bibinfo{year}{2008}).

\bibitem[{\citenamefont{McWilliams}(2010)}]{McWilliams:2010iq}
\bibinfo{author}{\bibfnamefont{S.~T.} \bibnamefont{McWilliams}}
  (\bibinfo{year}{2010}), \eprint{1012.2872}.

\bibitem[{\citenamefont{Vecchio}(2004)}]{Vecchio:2003tn}
\bibinfo{author}{\bibfnamefont{A.}~\bibnamefont{Vecchio}},
  \bibinfo{journal}{Phys. Rev.} \textbf{\bibinfo{volume}{D70}},
  \bibinfo{pages}{042001} (\bibinfo{year}{2004}), \eprint{astro-ph/0304051}.

\bibitem[{\citenamefont{Demorest et~al.}(2009)\citenamefont{Demorest, Lazio,
  Lommen, et.al.}}]{Demorest:2009ex}
\bibinfo{author}{\bibfnamefont{P.}~\bibnamefont{Demorest}},
  \bibinfo{author}{\bibfnamefont{J.}~\bibnamefont{Lazio}},
  \bibinfo{author}{\bibfnamefont{A.}~\bibnamefont{Lommen}}, \bibnamefont{and}
  \bibinfo{author}{\bibfnamefont{NANOGRAV} \bibnamefont{collaboration}}
  (\bibinfo{year}{2009}), \eprint{0902.2968}.

\end{thebibliography}

%
%
\end{document}